# Immuno-Histo X-ray Phase Contrast Tomography:
# New 3D imaging technique for molecular tomography

A. Quarta[1*], A. Sanna[1*], N. Pieroni[2], B. Parodi[3], F. Palermo[2], I. Bukreeva[2], M. Fratini[2], L. Massimi[2], D. Simeone[1], X. Le Guével[4], A. Bravin[5], I. Viola[2], E. Quintiero[2], G. Gigli[1], N. Kerlero de Rosbo[3]* L. Sancey[4]* A. Cedola[2]*

[1]*Institute of Nanotechnology-CNR, Lecce, Italy* [2] *Institute of Nanotechnology- CNR, Rome, Italy* [3]*Department of Neurosciences, Rehabilitation, Ophthalmology and Maternal-Fetal Medicine (DINOGMI), University of Genoa, Genoa, Italy,* [4] *Institute for Advanced Biosciences, Université Grenoble Alpes, INSERM U1209, CNRS UMR 5309, Grenoble, France.* [5] *University Milano Bicocca, Department of Physics "G. Occhialini", Milano MI*

*Equal contribution

## ABSTRACT

The still unmet ability of following the fate of specific cells or molecules within the whole body would give an outstanding breakthrough in the comprehension of disease mechanisms and in the monitoring of therapeutic approaches. Our idea is to push forward the bio-nanotechnology to a level where it serves the most advanced 3D bio-medical imaging to provide a multi-scale imaging ranging from the whole organ down to the cellular level, enabling high-resolution visualization of disease-relevant cells within the whole disease-altered biological context.
We present here the first proof-of-concept of a novel tomography procedure, Immuno-Histo-X-ray Phase Contrast Tomography (XPCT) that combines cutting-edge XPCT, which provides detailed image of the whole organ, with molecular imaging at the cellular level, identifying the relevant cells via an immunohistochemistry-based approach. We combine metal nanoparticles and a single-domain antibody that target relevant cells.
Our results lay the foundation for a new generation of 3D-bio-medical X-ray imaging.

## INTRODUCTION

Devising therapies that can reverse/prevent central nervous system (CNS) damage in neurodegenerative diseases necessitates in-depth investigation of disease mechanisms both in whole CNS and at cellular level. This, in turn, requires tools enabling visualization of disease-relevant vascular and neuronal networks (VN and NN) in the whole organ, together with affected cells, such as CNS-resident astroglia and infiltrating inflammatory immune cells, all related to neuroinflammation associated with neurodegeneration. The tools available today do not meet these requirements. Although jumping from 2D to 3D represented an outstanding breakthrough in the quality of imaging and information obtained, magnetic resonance and conventional X-ray-computed tomography fail to provide a satisfactory answer to the unmet medical-imaging needs for these diseases. Even more advanced emerging technology, such as the otherwise powerful X-ray phase contrast tomography (XPCT) [1,2,3,4,5,6,7,8], cannot simultaneously visualize disease-relevant VN and NN and identify affected cells unequivocally. Indeed, the identity of the different cells can only be speculated upon on a morphological/anatomical basis, and the lack of simultaneous cell authentication with 3D-imaging is a serious drawback of 3D tomography, XPCT included.

In this work, we completely change the concept of X-ray tomography from a mere 3D morphological investigation tool to a technique that provides an overview of the whole biomedical environment where the tiniest elements are resolved and selected cells can be monitored telling us their role in the disease and upon therapy. We propose here to combine, in one unique technique, XPCT whereby whole networks are displayed, with molecular imaging, whereby specific cells are identified via an immunohistochemistry-based approach with metal-tagged antibodies.

Structural imaging techniques were recently greatly improved by advances in nanotechnology. In particular, the use of nanoparticles (NP)-based targeted molecular probes, able to cross the blood brain barrier (BBB) and accumulate in brain parenchyma [9], has boosted the development of new imaging agents for assessing the brain function and CNS disorders. This is due to their unique features, such as size, high surface/volume ratio and surface chemistry, which can be conveniently manipulated to bind selected biomolecules. Colloidal NPs based on heavy elements, such as gold (Au) and gadolinium (Gd), are characterized by X-ray-attenuation properties and have been proposed and tested as biocompatible and effective long-circulating contrast agents for X-ray imaging and conventional computed tomography (CT) [9,10,11]. Although the use of NP-based targeted molecular probes enlarges the potential of standard imaging techniques, the limited spatial resolution and sensitivity of these techniques significantly limits their applicability when 3D imaging of neuronal and vascular networks of entire mouse CNS at cellular level is required. Identification of cells in tissue requires an immunohistochemistry-type approach, whereby cells can be authenticated through immunostaining of specific marker molecules. While monoclonal antibodies that target such specific markers have generally been immune reagents of choice, their usefulness for imaging is limited by their large size (> 150 kDa), which can lead to steric hindrance but also impedes their crossing of the BBB when imaging of the CNS is required. In contrast, single-domain antibodies (sdAbs) are ideally suited for imaging due to their small size enabling them to rapidly target antigenic epitopes at locations not easily accessible to conventional antibodies [13,14]. sdAbs are composed of the single heavy chain of antibodies that are expressed uniquely in camelids and lack the light chain. The capability to bind antigen is retained by a 15 kDa molecule that is 10 times smaller than a full Ab. In contrast to conventional Abs, sdAbs can cross the BBB, not only because of their small size, which allows extensive diffusion within cerebral tissue, but also because they could be genetically engineered to present a basic pI, which increases their BBB-crossing propensity. Such features, together with their stability and rapid clearance from blood, makes sdAbs best candidate for molecular imaging in brain. Indeed, although much of the research on sdAbs has focused on their potential role in therapeutic approaches, such as their ability to deliver attached cargo, e.g. in brain due to their BBB-crossing propensity, [15] there has been significant interest in their use for molecular imaging, through radionuclide-based, optical, and ultrasound imaging modalities, albeit with the major challenge of achieving a high-contrast signal of the targeting antibodies. At this stage, however, sdAbs have never been used in conjugation with metallic NPs for imaging by XPCT. Metallic NPs are highly versatile and tunable, in terms of protecting ligand engraftment, structure, and size (2-nm for the smallest); in particular, their very small size facilitates their elimination through the kidney when used alone [17,18].

We provide here the proof of concept for Immuno-Histo-XPCT (IH-XPCT) which boosts XPCT to a higher level equipping it with nanotechnology solutions enabling a novel immunohisto-X-ray tomography. To validate HI-XPCT, we have combined the high-resolution power of XPCT with in-vivo administration of innovative complexes of glial fibrillary acidic protein (GFAP) sdAb bearing Au-NPs, to identify astrocytes ex vivo in brain tissue.

## RESULTS

The surface of the ultrasmall Gold (Au) NPs was first passivated with a PEG shell and then coated with Nα,Nα-Bis(carboxymethyl)-L-lysine. The Ni coordination chemistry was used to bind the anti GFAP nanobodies, bearing a histidine tag. The average number of nanobodies linked per nanoparticle was kept close to 1 in order to associate each nanoparticle to a single bio-targeting unit. The final nanoconjugate was henceforth referred as Nano2.

The binding specificity of Nano2 NP to its biological target, the cytoplasmic protein GFAP was extensively evaluated.

To validate our concept, i.e. that specific cells could be immuno-monitored through XPCT, we administered the Nano2 *in vivo* in mice and imaged the tissue *ex vivo* by XPCT. The correct targeting of the astrocytes by the Nano2 was validated by confocal fluorescence microscopy to detect the Au concomitantly with the immunofluorescence emitted by a fluorophore-conjugated to an anti-astrocyte monoclonal antibody. To ensure that the blood-brain barrier would not impede the crossing of the Nano2 into the brain, we first tested the Nano2 in a mouse model of multiple sclerosis, the experimental autoimmune encephalomyelitis (EAE), in which inflammation increases the permeability of the BBB.

Experimentally, either unconjugated AuNPs or Nano2 were administered in-vivo by i) intranasal and ii) intravenous routes. Table 1 summarizes the samples analyzed.

Table 1: Different group of animals used in the XPCT analysis.

| | GROUP A | GROUP B | GROUP C | GROUP D | GROUP E | GROUP F Positive control | GROUP G Negative control |
|---|---|---|---|---|---|---|---|
| | EAE mice *intranasal administration of Au NPs* | EAE mice *intranasal administration of Nano2* | EAE mice *intravenous administration of Au NPs* | EAE mice *intravenous administration of Nano2* | Naïve mice *intranasal administration of Nano2* | Naïve mice *local* injection of Au NPs | Naïve mice *without injection* |

The ultimate goal was to evaluate the ability of NPs to reach the central nervous system and, possibly, accumulate in specific regions where astrocytes are present. Brains isolated from the euthanized mice were dissected into two parts: one hemisphere was imaged with different XPCT techniques; the second ones was analyzed by confocal microscopy (Figure 1a-c). The fluorescence of the AuNPs is mapped in red on sections in which cell nuclei have been marked in blue.

In order verify the affinity capacity of the Nano2, we performed additional confocal experiment on the mice brain of GROUP B, i.e., with intranasal administration of Nano2, using anti-Glast sdAbs to mark the astrocytes, as reported in Figure 1d. Nano2 were detected in the brain sections due to the presence of the AuNPs which are fluorescent in the near infrared optical region after blue light excitation. Some colocalizations between Nano2 and GLAST-expressing cells were observed after intranasal administration of Nano2, thus indicating the possibility to specifically target the astrocytes.

The obtained results can be summarized as follows:

- Nano2 are detected using dedicated microscopy settings
- Anti-Glast labeling is possible
- Colocalization of Nano2 and anti-Glast is clearly evidenced

The results on sample A and B, reported in Figure 1a-c, can be summarized in three points:
- The comparison between Figure 1a and 1b shows that the greatest accumulation of Nano2 and NPs in the brain occurs respectively when the injection is performed intranasally.
- The comparison between Figure 1a and 1c suggests that following intranasal injection Nano2 displays higher accumulation in the brain than unconjugated NPs.
- From Figure 1a and 1c, we can conclude that following intranasal injection, Nano2 are likely internalized into specific cells.

Sample E shows results similar to the sample B of figure 1a.

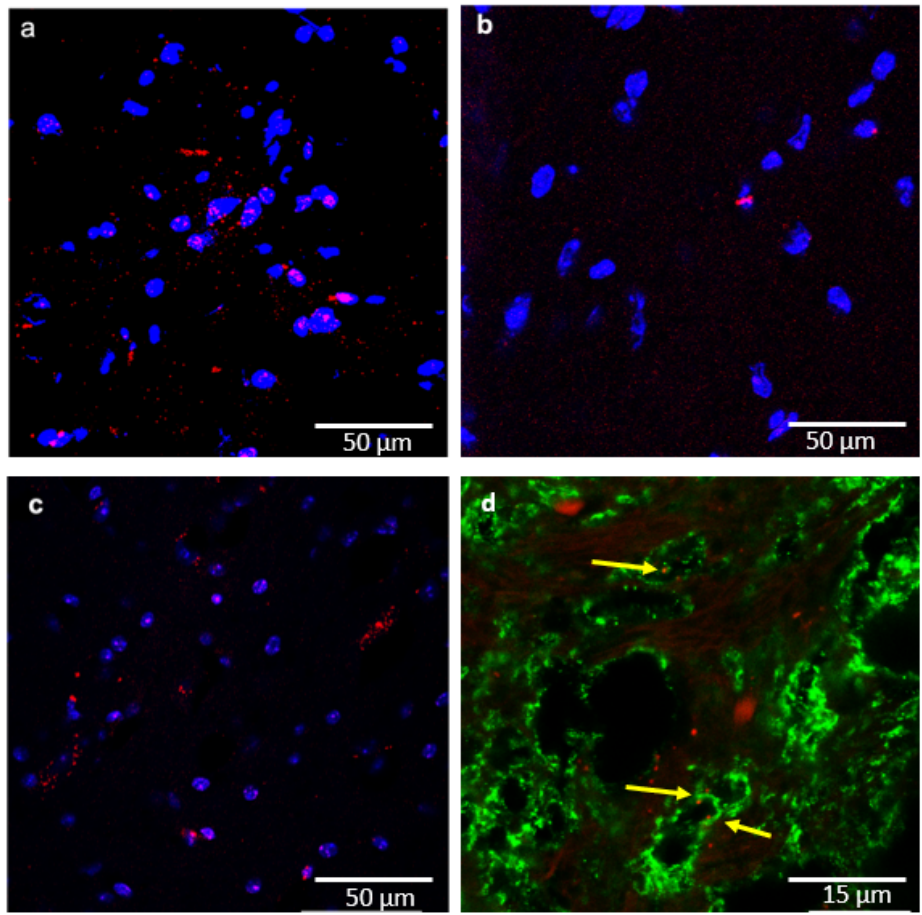

Figure 1: Confocal imaging of brain sections from mice of the different groups. a) Nano2 intranasally; b) NPs intranasally; c) Nano2 intravenously; d) Anti-Glast immunolabeling in presence of Nano2 with intranasal administration.

*NPs detectability*

The validation of the new X-ray imaging technique proposed here, IH-XPCT, involves XPCT measurement on mouse brain samples, in particular on the counterpart hemibrains not analyzed with confocal microscopy. XPCT drastically increases the image contrast of the soft tissues, classically considered invisible to standard X-ray tomography. Indeed, it combines the contrast generated by the variations of absorption coefficient in the sample with the contrast derived by the X-ray phase variations induced by the sample. Several experimental approaches have been developed to detect XPCT. In this work we combine the measurement from i) single-distance XPCT experiments, to image the whole hemibrain with high spatial resolution (0.7 microns); ii) XPC imaging experiment carried out above and below the gold (Au) L-absorption-edge (11.9 keV) to highlight the Au NPs with respect to the surrounding tissues and iii) XPCT experiment performed above and below the gold K-edge.

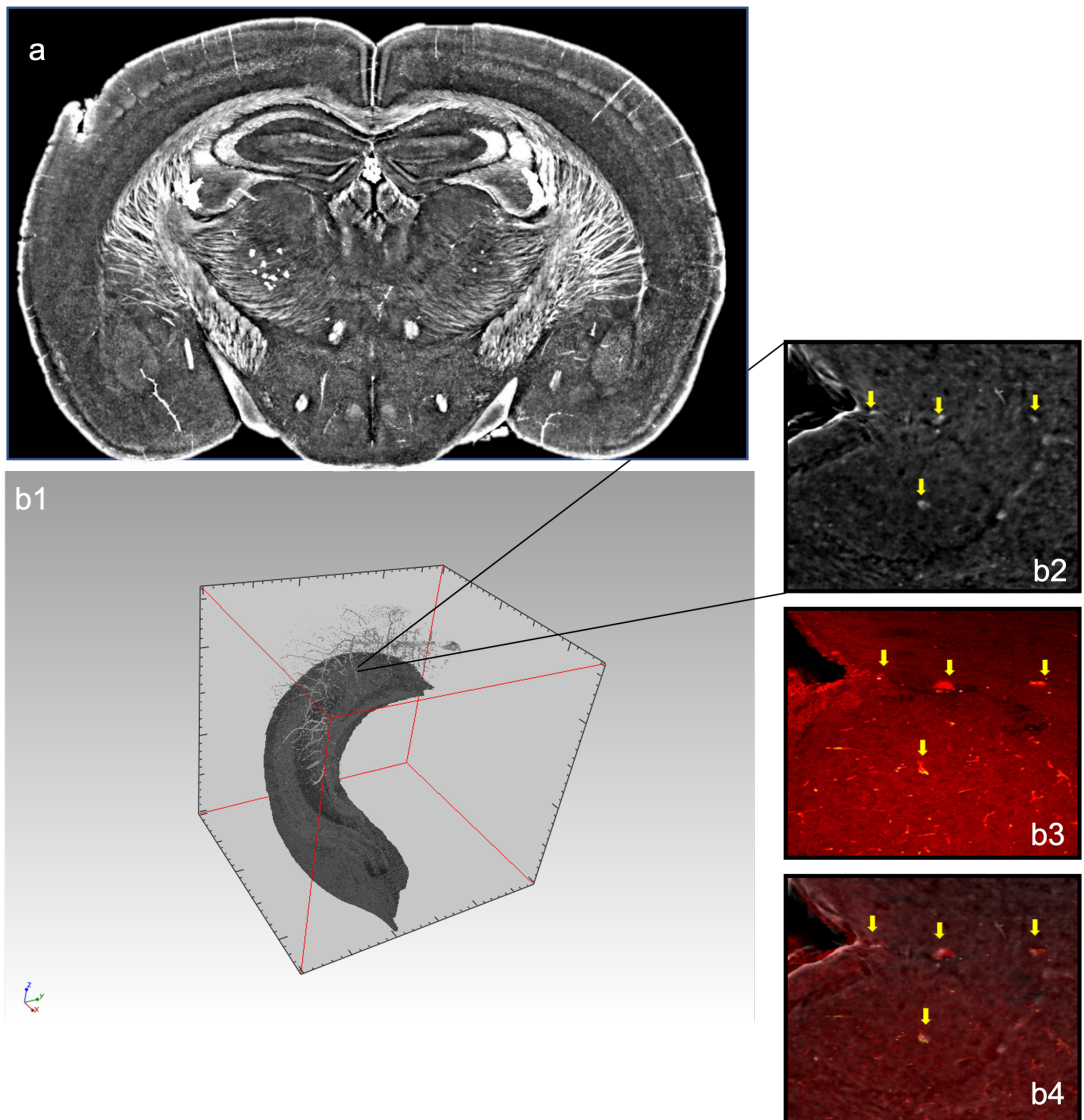

*Figure 2: a) 3D XPCT image of sample b1) zoom of the hippocampus region where Nano2 are supposed to be accumulated, following the confocal microscopy results; b2) X-ray phase contrast radiography, above and below L-edge of Au. The arrows indicate possible Nano2 accumulations; b3) Confocal microscopy of the same slice in figure b2). The arrows indicate Au florescence; b3) Co-registration of b2) and b3) where the arrows indicated the point of accumulation of possible Nano2.*

As shown in Figure 2a, XPCT provides a detailed characterization of the brain tissue; in particular, the vessels and the cells are very well mapped in white contrast (i.e., high electron density tissue). Following the confocal microscopy results, Nano2 are supposed to be accumulated in the hippocampus region. Figure 2.b1 shows a 3D reconstruction of the hippocampus of one sample A. To ascertain the Nano2 distribution in the hippocampus area, we performed X-ray phase contrast radiography in this brain region, above and below L-edge of Au. With this imaging approach we highlighted in the image the signal from Au with respect to all the other details. Figure 2.b2 shows the results obtained where the yellow arrows indicate white spots of signal from Au and therefore compatible with clusters of Nano2. Figure 2.b3 reports the result of confocal microscopy of the same slab as in figure 2.b2. The arrows indicate the spots where Au florescence is high. The co-registration of the X-ray imaging and the Au fluorescence, shown in Figure 2.b4, demonstrates the specific accumulation of Au Nano2 clusters, as indicated by the arrows in Figure 2.b4.

It is well known [18, 19] that after an intranasal injection a certain amount of NPs lies along the blood vessels. Segmentation from the XPCT images shows a clear difference between the vessels of the negative control (non-injected) sample and those of the sample injected with Nano2. The former shows a clear border between the surrounding tissue and the lumen (black, low absorbing) filled with white (high absorbing) inhomogeneous distribution from blood clots. On the other hand, the vessels of the sample injected with Nano2 appear with a white halo at the interface with the lumen (Figure 3b). This halo is compatible with the presence of Nano2 lying along the vessels, as expected. A comparison with the confocal microscopy images of the vessels supports our hypothesis: as shown in Figure 3c the Au fluorescence signal (yellow in the color scale) looks just outside the border of the lumen in the injected sample, while the images of the non-injected sample (Figure 3d) show the same characteristics as in the XPCT images.

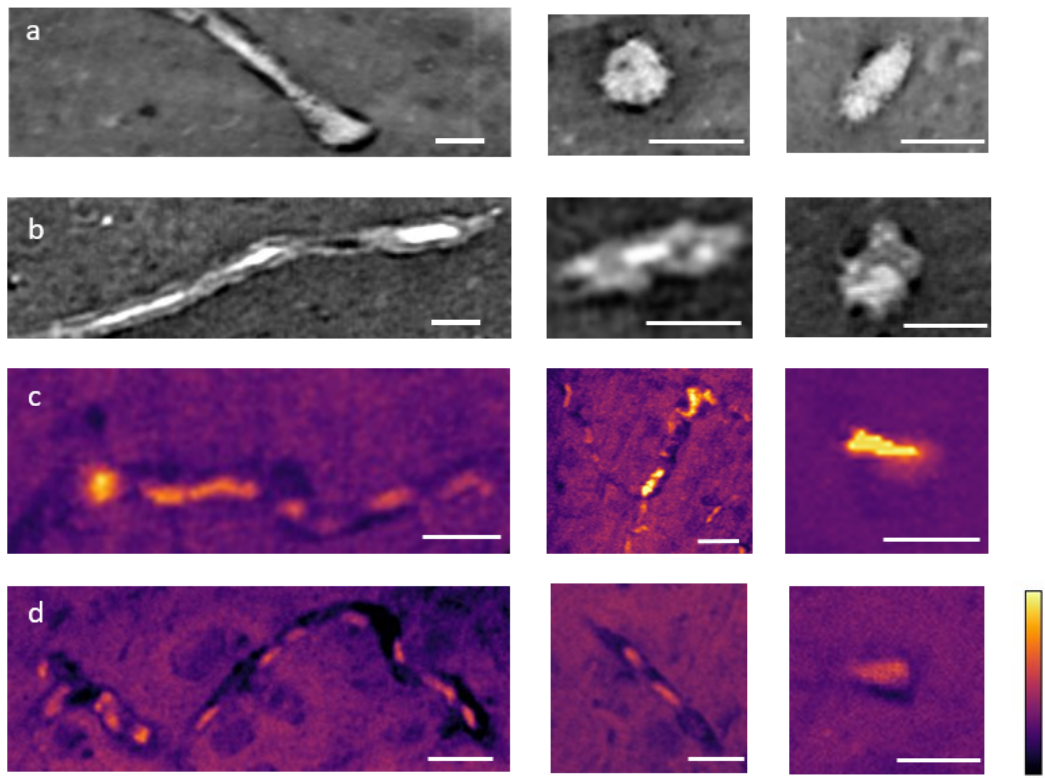

*Figure 3: Vessel sections. a. XPCT images of vessel sections of negative control mouse. b. XPCT images of vessel sections of a mouse from GROUP B. c. Confocal microscopy images of vessel sections of a mouse from GROUP B. d. Confocal microscopy images of vessel sections of negative control mouse Scale bar: 30 µm.*

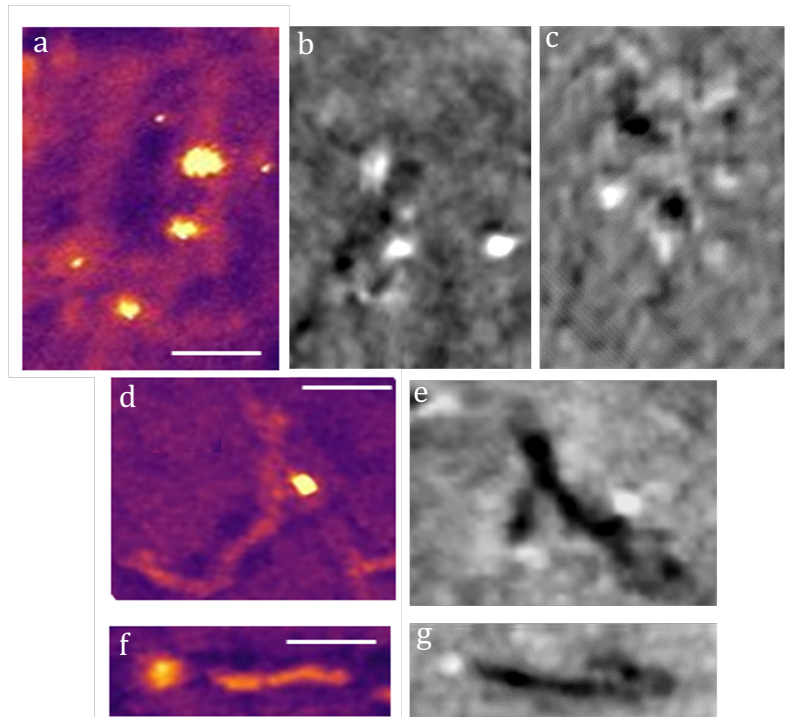

*Figure 4: Vessels sections of a mouse from GROUP B. a, d, f. Confocal microscopy images. b, c, e, g. Tomographic Au-K-edge Subtraction images. Scale bar: 30 µm.*

**CONCLUSIONS**

X-ray phase contrast tomography is a non-destructive imaging technique which enables multi-scale 3D biomedical imaging of neuronal and vascular networks with a field of view ranging from the single cell through the brain as a whole, without the need of slicing or pre-processing of the tissues; 3D "virtual histology" of the sample is a major achievement of XPCT. However, a crucial challenge for this technique is the ability to univocally identify the cells of interest. The new approach proposed here, Immuno-Histo X-Ray Phase Contrast Tomography, addresses this issue by pushing forward the bio-nanotechnology to enrich the otherwise *mere morphological* 3D tomography with *tissue-specificity*. The application of this new technique is of great interest in the context of preclinical studies on neurodegenerative diseases, in which it is essential to detect and investigate specific neuropathological signatures and to follow-up the effects of therapies at cellular level.

In this work, we demonstrate that metal-NPs combined with sdAbs, Nano2, cross the BBB, reach the CNS and the cells of interest and finally XPCT provides 3D image of these Nano2 systems within the brain tissue.

The next step of our work will be to use NPs of different metals, each coupled to different sdAbs in order to tag and identify different cell types in a single animal.

## ACKNOWLEDGMENTS

The authors are grateful to Dr. Lafaye from Institut Pasteur for supplying the anti GFAP nanobody.
This work was supported by Regione Puglia and CNR for Tecnopolo per la Medicina di Precisione. D.G.R. n. 2117 of 21.11.2018.

## References

[1] *Exploring Alzheimer's disease mouse brain through X-ray phase contrast tomography: From the cell to the organ,* Massimi, L., Bukreeva, I., Santamaria, G., Fratini, M., Corbelli, A., Brun, F., Fumagalli, S., Maugeri, L., Pacureanu, A., Cloetens, P., Pieroni, N., Fiordaliso, F., Forloni, G., Uccelli, A., Kerlero de Rosbo, N., Balducci, C., Cedola, A., (2019) NeuroImage, 184, pp. 490-495. DOI: 10.1016/j.neuroimage.2018.09.044

[2] *Simultaneous submicrometric 3D imaging of the micro-vascular network and the neuronal system in a mouse spinal cord,* Fratini, M., Bukreeva, I., Campi, G., Brun, F., Tromba, G., Modregger, P., Bucci, D., Battaglia, G., Spanò, R., Mastrogiacomo, M., Requardt, H., Giove, F., Bravin, A., Cedola, A., (2014) Scientific Reports, 5, art. no. 8514, .DOI: 10.1038/srep08514

[3] *Quantitative 3D investigation of Neuronal network in mouse spinal cord model* Bukreeva, I., Campi, G., Fratini, M., Spanò, R., Bucci, D., Battaglia, G., Giove, F., Bravin, A., Uccelli, A., Venturi, C., Mastrogiacomo, M., Cedola, A., (2017) Scientific Reports, 7, art. no. 41054., DOI: 10.1038/srep41054

[4] *X-Ray Phase Contrast Tomography Reveals Early Vascular Alterations and Neuronal Loss in a Multiple Sclerosis Model*, Cedola, A., Bravin, A., Bukreeva, I., Fratini, M., Pacureanu, A., Mittone, A., Massimi, L., Cloetens, P., Coan, P., Campi, G., Spanò, R., Brun, F., Grigoryev, V., Petrosino, V., Venturi, C., Mastrogiacomo, M., Kerlero De Rosbo, N., Uccelli, A., (2017) Scientific Reports, 7 (1), art. no. 5890., DOI: 10.1038/s41598-017-06251-7

[5] *X-ray Phase Contrast Tomography Serves Preclinical Investigation of Neurodegenerative Diseases*, Palermo et al., Front. Neurosci., 09 November 2020 |https://doi.org/10.3389/fnins.2020.584161

[6] Three dimensional visualization of engineered bone and soft tissue by combined x-ray micro- diffraction and phase contrast tomography, A Cedola, G Campi, D Pelliccia, I Bukreeva, **M** Fratini, et al. Physics in Medicine & Biology 59 (1), 189

[7] Engineered bone from bone marrow stromal cells: a structural study by an advanced x-ray microdiffraction technique, A Cedola, M Mastrogiacomo, M Burghammer, V Komlev, P et al, Physics in Medicine & Biology 51 (6), N109.

[8] A three-image algorithm for hard x-ray grating interferometry, D Pelliccia, L Rigon, F Arfelli, RH Menk, I Bukreeva, A Cedola, Optics express 21 (16), 19401-194

[9] *Nanoparticles enhance brain delivery of blood–brain barrier-impermeable probes for in vivo optical and magnetic resonance imaging*, Koffie et al., PNAS 2011, 108, 46: 18837-18842, https://doi.org/10.1073/pnas.1111405108

[10] *Single-cell resolution in Single-high resolution synchrotron Xray CT imaging with gold nanoparticles.* Schültke E, Menk R, Pinzer B, Astolfo A, Stampanoni M, Arfelli F, Harsan LA, Nikkhah G.J Synchrotron Radiat. 2014 Jan;21(Pt 1):242-50. doi: 10.1107/S1600577513029007.

[11] *PEG-modified gadolinium nanoparticles as contrast agents for in vivo micro-CT.*
Cruje C, Dunmore-Buyze PJ, Grolman E, Holdsworth DW, Gillies ER, Drangova M.Sci Rep. 2021 Aug 16;11(1):16603.

[12] *Ultrasmall Silica-Based Bismuth Gadolinium Nanoparticles for Dual Magnetic Resonance-Computed Tomography Image Guided Radiation Therapy.*
Detappe A, Thomas E, Tibbitt MW, Kunjachan S, Zavidij O, Parnandi N, Reznichenko E, Lux F, Tillement O, Berbeco R.Nano Lett. 2017 Mar 8;17(3):1733-1740.

[13] *Single-Domain Antibodies as Therapeutic and Imaging Agents for the Treatment of CNS Diseases.,* Bélanger K, Iqbal U, Tanha J, MacKenzie R, Moreno M, Stanimirovic D.Antibodies (Basel). 2019 Apr 5;8(2):27. doi: 10.3390/antib8020027.

[14] *Camelid single-domain antibodies: A versatile tool for in vivo imaging of extracellular and intracellular brain targets.,* Li T, Vandesquille M, Koukouli F, Dudeffant C, Youssef I, Lenormand P, Ganneau C, Maskos U, Czech C, Grueninger F, Duyckaerts C, Dhenain M, Bay S, Delatour B, Lafaye P.J Control Release. 2016 Dec 10;243:1-10. doi:10.1016/j.jconrel.2016.09.019.

[15] *Single domain antibody-based vectors in the delivery of biologics across the blood-brain barrier: a review.* Gao Y, Zhu J, Lu H.Drug Deliv Transl Res. 2021 Oct;11(5):1818-1828.

[16] *Surface functionalization of gold nanoclusters with arginine: a trade-off between microtumor uptake and radiotherapy enhancement.*
Broekgaarden M , Bulin AL , Porret E , Musnier B , Chovelon B , Ravelet C , Sancey L , Elleaume H , Hainaut P , Coll JL , Le Guével X .Nanoscale. 2020 Apr 3;12(13):6959-6963.

[17] *Long-term in vivo clearance of gadolinium-based AGuIX nanoparticles and their biocompatibility after systemic injection.*
Sancey L, Kotb S, Truillet C, Appaix F, Marais A, Thomas E, van der Sanden B, Klein JP, Laurent B, Cottier M, Antoine R, Dugourd P, Panczer G, Lux F, Perriat P,

Motto-Ros V, Tillement O.ACS Nano. 2015 Mar 24;9(3):2477-88. doi: 10.1021/acsnano.5b00552.

[18] *Intranasal delivery to the central nervous system: mechanisms and experimental considerations.* Dhuria SV, Hanson LR, Frey WH 2nd., J Pharm Sci. 2010 Apr;99(4):1654-73. doi: 10.1002/jps.21924. PMID: 19877171.

[19] *Intranasal administration of CNS therapeutics to awake mice.* Hanson LR, Fine JM, Svitak AL, Faltesek KA, J Vis Exp. 2013 Apr 8;(74):4440. doi: 10.3791/4440. PMID: 23608783; PMCID: PMC3653240.